\def\tit#1{``#1,''}
\long\def\comment#1{}

\let\ox=\otimes

\newcommand{\Mx}[3]{\left#1\begin{array}{rrrr}#2\end{array}\right#3}

\newcommand{\ket}[1]{| #1 \rangle}
\newcommand{\bra}[1]{\langle #1 |}

\newcommand{\vxv}[1]{\ket{#1}\bra{#1}}
\newcommand{\brkt}[2]{\langle #1 \mid #2 \rangle}
\newcommand\Tr{\mathop\mathrm{Tr}}
\newcommand\card{\mathop\mathrm{card}}

\newcommand\C{\mathbb C}
\newcommand\R{\mathbb R}
\newcommand\Z{\mathbb Z}
\newcommand\N{\mathbb N}
\newcommand\U{\mathsf U}
\newcommand\iint{\int\!\!\int}
\newcommand{\wave}[1]{\hbox to #1{\leaders\hbox{${\sim}\!$}\hfil}}
\newcommand{\rightwave}{\mathop{\wave{2cm}\!{\to}}\limits}

\newcommand{\GE}{\geqslant}
\newcommand{\LE}{\leqslant}
\newcommand{\op}[1]{\boldsymbol{#1}}

\newcommand\eq[1]{Eq.~(\ref{#1})}
\newcommand\Sec[1]{Sec.~\ref{sec:#1}}
\documentclass{article}
\usepackage{exscale}
\usepackage{amssymb,amsbsy}
\begin{document}
\begin{titlepage}
\def\newpage{\relax}
\noindent{$_{}$\huge\hrulefill %
\smash{I\kern-.16em B\kern -.05em r{\Large\kern-.09em i\kern-.09em}ef Report}}
\noindent\hrule
\title{\bf {\huge {\sf ALEPH--QP} (\boldmath{$\aleph_{QP}$)}:} \\
Universal hybrid quantum processors\\
\large with continuous and discrete quantum variables}
\author{{\em Alexander Yu.\ Vlasov%
 \thanks{Electronic mail: \tt Alexander.Vlasov@PObox.spbu.ru}}\\
{\small FRC/IRH, 197101 Mira Street 8, St.--Petersburg, Russia}}
\date{}
\maketitle
\begin{abstract}
\noindent
A quantum processor (the programmable gate array) is a quantum network with a
fixed structure. A space of states is represented as tensor product of data
and program registers. Different unitary operations with the data register
correspond to ``loaded'' programs without any changing or ``tuning'' of
network itself. Due to such property and undesirability of entanglement
between program and data registers, universality of quantum processors is
subject of rather strong restrictions. By different authors was developed
universal ``stochastic'' quantum gate arrays. It was proved also, that
``deterministic'' quantum processors with finite-dimensional space of states
may be universal only in approximate sense. In present paper is shown, that
using hybrid system with continuous and discrete quantum variables, it is
possible to suggest a design of strictly universal quantum processors.
It is shown also that ``deterministic'' limit of specific programmable
``stochastic'' $U(1)$ gates (probability of success becomes unit for
infinite program register), discussed by other authors, may be essentially
same kind of hybrid quantum systems used here.
\end{abstract}
\end{titlepage}

\section{Introduction}
\label{sec:intro}

The quantum programmable gate array \cite{NC97,VC00,VMC01,Vla01a} or
{\em quantum processor} \cite{Vla01b,HBZ01} --- is a quantum
circuit with fixed structure.
Similarly with usual processor here are {\em data register} $\ket{D}$ and
{\em program register} $\ket{P}$. Different operations $\op u$ with data are
governed by a state of the program, {\em i.e.}\ it may be described as
\begin{equation}
\U \colon \bigl(\ket{P} \ox \ket{D}\bigr) \mapsto \ket{P'} \ox (\op u_P\ket{D}).
\label{qproc}
\end{equation}
Each register --- is a quantum system\footnote{usually finite-dimensional}
and may be represented for particular task using
qubits \cite{NC97,VC00,VMC01,Vla01a}, qudits \cite{Vla01b,HBZ01}, {\em etc.}.

It can be simply found \cite{NC97}, that \eq{qproc} is compatible with
unitary quantum evolution, if different states of {\em program register}
are orthogonal --- due to such requirement number of accessible programs
coincides with dimension of Hilbert space and it produces some challenge for
construction of universal quantum processors. It was suggested few ways
around such a problem: to use specific ``stochastic'' design of universal
quantum processor \cite{NC97,VC00,VMC01,HBZ01},
to construct (non-stochastic) quantum processor with possibility to
approximate any gate with given precision \cite{VC00,VMC01,Vla01a,Vla01b}
(it is also traditional approach to universality \cite{Deu85,Deu89,Ek95},
sometime called ``universality in approximate sense'' \cite{Cle99}).

Here is discussed an alternative approach for strictly universal quantum processor
--- to use continuous quantum variables in program register and discrete ones
for data, {\em i.e.\ hybrid} quantum computer \cite{Llo00}. In such a case
number of different programs is infinite and it provides possibility to
construct strictly universal hybrid quantum processor for initial
(``deterministic'') design described by \eq{qproc}. It is enough to
provide procedures for one-qubit rotations with three real parameters
together with some finite number of two-gates \cite{Cle99,Gate95}.

It is shown also, that hybrid quantum gates used in this article can be
considered not only as limit of deterministic design \cite{Vla01a,Vla01b},
but also coincide with discussed in \cite{VMC01} ``deterministic limit'' of
a special case of programmable $U(1)$ ``stochastic'' gates with probability
of fail tends to zero for infinite program register.

\section{Construction of hybrid quantum processors}\label{sec:hybproc}

In finite-dimensional case unitary operator $\U$ satisfying \eq{qproc}
can be simply found \cite{Vla01a,Vla01b}. Let us consider case with
$\ket{P'} = \ket{P}$ in \eq{qproc}. It was already mentioned,
that states $\ket{P}$ of program register corresponding to different
operators $\op u_P$ are orthogonal and, so, may be chosen as basis. In such
a basis $\op u_P$ is simply set of matrices numbered by integer index $P$,
and operator $\U$ \eq{qproc} can be written as block-diagonal
$NM \times NM$ matrix:
\begin{equation}
\U = \Mx({\op u_1\\&\op u_2&&\smash{\mbox{\Huge$0$}}\\
                 &&\ddots\\\smash{\mbox{\Huge$0$}}&&&\op u_M}),
\end{equation}
with $N \times N$ matrices $\op u_P$,
if dimensions of program and data registers are $M$ and $N$ respectively;
\begin{equation}
\U = \sum_{P=1}^M \ket{P}\bra{P} \ox \op u_P,
\label{conddyn}
\end{equation}
It is {\em conditional quantum dynamics} \cite{Joz95}.
For quantum computations with qubits $M=2^m$, $N=2^n$.

Generalization to hybrid system with program register described
by one continuous quantum variable and qubit data register is straightforward.
The states of program register may be described as Hilbert space of functions
on line $\psi(x)$. In coordinate representation a basis is
\begin{equation}
 \ket{q} = \delta(x-q); \quad \brkt{q}{\psi(x)} = \psi(q).
\label{q}
\end{equation}

To represent some continuous family of gates $\op u_{(q)}$ acting on data
state, say phase rotations
\begin{equation}
 \op\theta_{(q)}=\exp(2 \pi i q \op\sigma_3),
\label{thetaq}
\end{equation}
it is possible
to write continuous analog of \eq{conddyn}:
\begin{eqnarray}
 &\U = \int dq \bigl(\ket{q}\bra{q} \ox \op u_{(q)}\bigr) , \\
 &\U\bigl(\psi(x)\ket{s}\bigr) = \int \delta(x-q) \psi(q) \ket{\op u_{(q)}s} dq=
 \psi(x)\ket{\op u_{(x)}s},
\end{eqnarray}
where $\ox$ is omitted because $\ket{\psi}\ket{s}$
can be considered as product of scalar function $\psi(x)$ on complex vector
$\ket{s}$.
Finally:
\begin{equation}
 \U (\ket{q}\ket{s}) = \ket{q}\ket{\op u_{(q)}s}.
\label{ctrQ}
\end{equation}

It is convenient also to use momentum basis, {\em i.e.}:
\begin{equation}
 \ket{\tilde p} = e^{i p x}; \quad
 \brkt{\tilde p}{\psi(x)} = {\textstyle\int} e^{-i p x} \psi(x) dx
 \equiv \tilde\psi(p).
\end{equation}
(where $\tilde\psi$ is Fourier transform of $\psi$)
and operator $\tilde\U$:
\begin{eqnarray}
 &\tilde\U = \int dp \bigl(\vxv{\tilde p} \ox \op u_{(p)}\bigr),& \\
 &\tilde\U\bigl(\psi(x)\ket{s}\bigr) =
 \int e^{i p x} \bigl(\int e^{-i p x'} \psi(x') dx'\bigr) \ket{\op u_{(p)}s} dp= &
\nonumber \\
 &= \iint e^{i p (x-x')} \psi(x') \ket{\op u_{(p)}s} dx' dp
 = -\iint e^{-i p q} \psi(x-q) \ket{\op u_{(p)}s} dq dp.&
\end{eqnarray}
Here $\tilde\U$ is not rewriting $\U$ in momentum basis, it is other
operator with property:
\begin{equation}
 \tilde\U (\ket{\tilde p}\ket{s}) = \ket{\tilde p}\ket{\op u_{(p)}s}.
\label{ctrP}
\end{equation}

It has simpler physical interpretation. Let us consider scattering of some
scalar particle on quantum system with two states (qubit). Then
\eq{ctrP} can be written symbolically as:
\[
 {\rightwave^{\ket{\exp(i k x)}_1}}\ %
 {\mathop{{\bullet}\mkern-13mu{\nearrow}}\limits^{\ \ket{s}_2}}
 \quad \Longrightarrow \quad
 {\mathop{{\bullet}\mkern-13.5mu{\nwarrow}}\limits^{\op u_{(k)}\ket{s}_2}}\ %
 {\rightwave^{\ket{\exp(i k x)}_1}}
\]

Using such approach with hybrid program register (few continuous variables
for different qubit rotations and discrete ones for two-gates like CNOT), it
is possible to suggest design of universal quantum processor with qubits data
register.

Hilbert space of hybrid system with $k$ continuous and $M=2^m$ discrete
quantum variables can be considered as space of $\C^M$--valued functions
with $k$ variables
$$
 F(x_1,\dots,x_k) \colon \R^k \to \C^M.
$$

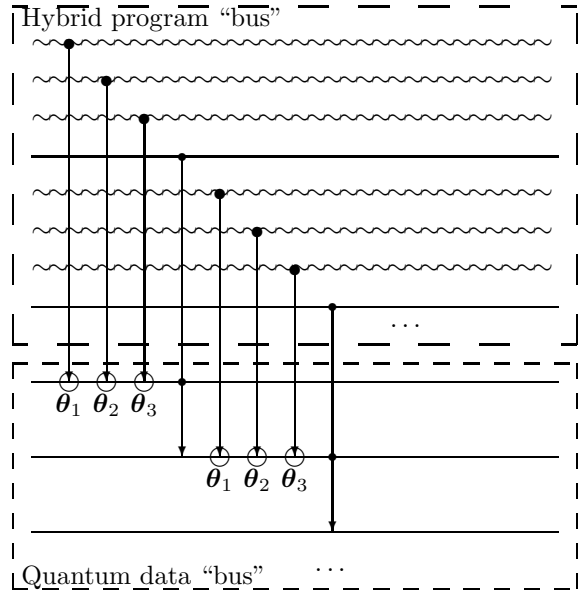
\begin{figure}[ht]
\begin{center}
\unitlength=1mm
\linethickness{0.4pt}
\begin{picture}(85,75)
\put(10,75){\makebox(0,0)[lc]{\wave{7cm}}}
\put(10,70){\makebox(0,0)[lc]{\wave{7cm}}}
\put(10,65){\makebox(0,0)[lc]{\wave{7cm}}}
\put(10,60){\line(1,0){70}}
\put(10,55){\makebox(0,0)[lc]{\wave{7cm}}}
\put(10,50){\makebox(0,0)[lc]{\wave{7cm}}}
\put(10,45){\makebox(0,0)[lc]{\wave{7cm}}}
\put(10,40){\line(1,0){70}}
\put(10,30){\line(1,0){70}}
\put(10,20){\line(1,0){70}}
\put(10,10){\line(1,0){70}}
\put(15,75){\vector(0,-1){45}}
\put(15,75){\circle*{1.5}}
\put(20,70){\vector(0,-1){40}}
\put(20,70){\circle*{1.5}}
\put(25,65){\circle*{1.5}}
\put(25,65){\vector(0,-1){35}}
\put(30,60){\vector(0,-1){40}}
\put(30,60){\circle*{1}}
\put(30,30){\circle*{1}}
\put(35,55){\vector(0,-1){35}}
\put(35,55){\circle*{1.5}}
\put(40,50){\vector(0,-1){30}}
\put(40,50){\circle*{1.5}}
\put(45,45){\vector(0,-1){25}}
\put(45,45){\circle*{1.5}}
\put(50,40){\vector(0,-1){30}}
\put(50,40){\circle*{1}}
\put(50,20){\circle*{1}}
\put(60,37.5){\makebox(0,0)[cc]{\ldots}}
\put(50,5){\makebox(0,0)[cc]{\ldots}}
\put(15,30){\circle{2.5}}
\put(15,27){\makebox(0,0)[cc]{$\op\theta_1$}}
\put(20,30){\circle{2.5}}
\put(20,27){\makebox(0,0)[cc]{$\op\theta_2$}}
\put(25,30){\circle{2.5}}
\put(25,27){\makebox(0,0)[cc]{$\op\theta_3$}}
\put(35,20){\circle{2.5}}
\put(35,17){\makebox(0,0)[cc]{$\op\theta_1$}}
\put(40,20){\circle{2.5}}
\put(40,17){\makebox(0,0)[cc]{$\op\theta_2$}}
\put(45,20){\circle{2.5}}
\put(45,17){\makebox(0,0)[cc]{$\op\theta_3$}}
\put(7.5,35){\dashbox{4}(75,45)[lt]{$\mathstrut$ Hybrid program ``bus''}}
\put(7.5,2.5){\dashbox{2}(75,30)[lb]{ Quantum data ``bus''}}
\end{picture}
\caption{Hybrid quantum circuit.
 ($\op\theta_k \equiv \op\theta_k(p) = e^{2 \pi i\, p \op\sigma_k}$)}
\label{fignet}
\end{center}
\end{figure}

For construction of universal processor it is possible to use three
continuous variables\footnote{Angular parametrization of $SU(2)$} for each
qubit together with discrete variables for control of two-qubit gates
(see Fig.~\ref{fignet}).

It should be mentioned, that it is rather simplified model. More rigor
consideration for different physical examples may include different
functional spaces, distributions, functions localized on discrete set
of points, and symbol ``$\smallint$'' or scalar product used in formulas
above in such a case should be defined with necessary care. Due to such
a problem in many works about quantum computations with continuous
variables is used Heisenberg approach and expressions with operators like
coordinate $\op Q$ and momentum $\op P$ \cite{LlBr98}.

Heisenberg approach maybe simplifies description, but hides some subtleties.
For example, in many models variables hardly could be called ``continuous'',
because they may be described as set of natural numbers, {\em i. e.}\ terms
``infinite,'' ``nonfinite'' maybe better for such quantum variables.

Let us consider example with qubit controlled by continuous variable
described above $\op u_{(q)}\equiv\op\theta_{(q)}$ \eq{thetaq}. In such a case
it is enough to use in operator $\U$ \eq{ctrQ} only interval of values
$0 < q \LE 2\pi$ or even consider Hilbert space of periodic functions
$\psi(q)$, like phases. But in such a case in dual space momenta have only
discrete set of values $p \in \Z$ and because both spaces connected by
Fourier transform, it is example of relation between periodical
functions of continuous variable and functions defined on infinite, but
discontinuous set $\Z$ of integer numbers.

Here is important issue: the commutation relations like $i[\op P,\op Q]=\op 1$
are not compatible with linear algebra of any finite matrices,%
\footnote{It is simple to show, taking trace of the commutator for
$D \times D$ matrices:
$i\Tr[\op P,\op Q]=i\Tr(\op{PQ})-i\Tr(\op{QP})=0 \ne \Tr(\op 1) = D$}
but may be simply satisfied by infinite-dimensional operator algebras, like
Schr\"odinger representation $\op Q=x$, $\op P = -i\,d/dx$. But here is yet another
problem --- integer and real numbers are used for representation of
infinite quantum variables, but {\em cardinality} of the sets are different,
$\card \N = \aleph_0$, $\card \R = \aleph$. To avoid discussion, related with
the cardinality issues, Russell paradox, {\em etc.}, here is used some formal
cardinality $\aleph_{QP}$ of ``quantum infinite variables'', {\em i.e.} any
model of infinite numbers appropriate for introduction of Heisenberg relations%
\footnote{So {\sf ALEPH--QP} --- is shortcut for ``quantum processor with
continuous or unlimited discrete variables.''}.

\bigskip

It should be mentioned, that term ``hybrid'' is used also with other
meaning \cite{hybr2}. Formally it is different thing, but for discussed
strategy for hybrid quantum processors, these two topics are close linked.
Let us discuss it briefly. For realistic design of quantum
computers, it is useful to have some language for joint description
with more convenient classical microdevices, which could be used as some
base for development of quantum processors. Generally such a task is
very difficult (if possible at all) and has variety of different approaches.

But there is especially simple idea, that could be applied for model under
consideration. The quantum gates and ``wires'' may be roughly treated as
(pseudo)classical, if only elements of computational basis are accepted in a
model as states of system and gates are also may not cause any superposition and
directly corresponds to set of invertible classical logical gates \cite{Ek95}.

Really, such a model is still quantum, but has closer relation with usual
classical circuits and so may reduce some difficulties in description of
hybrid classical--quantum processor design. It was already discussed in
\cite{Vla01a,Vla01b}, that from such point of view program register can be
treated as pseudo-classical\footnote{Really it was used even more specific design
with {\em intermediate register (bus)} between program and data one.}. It was
design with finite number of state in program register.

Similar procedure without difficulties may be extended for continuous case,
but now it corresponds to continuous classical variables, {\em i.e.}\ it is
similar either with analogue classical control or with more detailed description
of usual microprocessor, when inputs and outputs are described not as abstract
zeros and ones, but as real dynamically changed classical continuous signals
(fields, currents, laser beams, {\em etc.}).

\section{Comparison with limit of ``stochastic'' models}

In this paper was used design of universal hybrid quantum processor, that
could be considered as some limit of approximately universal ``deterministic''
quantum processors \cite{Vla01a,Vla01b}, when size of program register
formally becomes unlimited. On the other hand, in \cite{VC00,VMC01} is
considered design of programmed ``stochastic'' $U(1)$ gates with probability
of success becomes arbitrary close to unit with extension of program register
and so such design formally also becomes deterministic for infinite size
of program register.

Conceptually, the ``probabilistic,'' ``stochastic'' design of quantum
processors \cite{NC97,VC00,VMC01,HBZ01,HBZ02} is rather tricky question,
but it is not discussed here in details.

For our purposes is enough to use ``stochastic'' quantum circuit
\cite{VC00,VMC01} for application of gate $\op\theta_{(q)}$ \eq{thetaq} with
probability of success $p=1-1/M$ for size $M=2^m$ of ($m$-qubits) program
register with existing of ``deterministic'' limit $p=1$ for $M \to \infty$
\cite{VMC01}. Let us, without embarking in discussion about specific
problems of ``stochastic'' model, consider the limit and show,
that it is essentially same programmable phase gates discussed in
\Sec{hybproc}.

The construction is straightforward. For ``encoding transformations''
$\op\theta_{\alpha}$ to state of $m$-qubits program register in
\cite{VC00,VMC01} is used a family of states
\begin{equation}
 \ket{\Phi_{\alpha,m}} = \bigotimes_{k=0}^{m-1} \ket{\phi_{2^k\alpha}},
\mbox{ where }
\ket{\phi_a} \equiv \frac{1}{\sqrt{2}}(e^{i a/2}\ket{0} + e^{-i a/2}\ket{1}).
\label{Phi0}
\end{equation}
It can be rewritten as\footnote{Here is used ``inverted'' binary notation
for $\ket{K}$, $0 \LE K < M$, {\em i.e.}\ $\ket{b_0 b_1 b_2 \ldots }$
corresponds to $K = b_0+2 b_1 +2^2 b_2+ \cdots$. Anther choice is to save
standard binary notation, but to change order of terms in initial tensor
product \eq{Phi0} to opposite one.}
\begin{equation}
\ket{\Phi_{\alpha,m}} = \frac{e^{i\alpha(M-1)/2}}{\sqrt{M}}
\sum_{K=0}^{M-1} e^{-i K \alpha}\ket{K} \qquad (M = 2^m)
\label{ketPhi}
\end{equation}
and for $\alpha = - 2 \pi p / M$ with integer $p$ states \eq{ketPhi} coincide
with usual {\em momentum} basis $\ket{\tilde{p}}=\ket{\Phi_{-2\pi p/M,m}}$
$(p \in \Z,\ 0 \LE p < M)$ of $M$-dimensional Hilbert space.

Such elements $\ket{\tilde{p}}$ may be used as $M$ orthogonal basic states of
program register in ``deterministic'' quantum processor [{\em c.f.}
\eq{conddyn}],
\begin{equation}
\tilde\U = \sum_{p=0}^{M-1} \vxv{\tilde{p}} \ox \op\theta_{(2\pi p / M)}.
\end{equation}

The ``deterministic'' approach uses only the computational basis $\tilde{p}$
and it prevents from entanlement between program and data registers. Stochastic
$U(1)$ approach \cite{VC00,VMC01} uses $\ket{\Phi_{\alpha,m}}$ with arbitrary
$\alpha$ and for finite-dimensional case such states are not always orthogonal,
but here is possible to do not discuss the issues related with interpretation
of {\em quantum measurements} used for ``probabilistic'' calculations of
$\op\theta_{(\alpha)}$ for entangled case $\alpha \ne 2 \pi p / M$, because
for infinite-dimensional case {\em all} states $\ket{\Phi_{\alpha,\infty}}$
are orthogonal.

So, continuous (infinite) limit of ``stochastic'' $U(1)$ programmable gates
suggested in \cite{VC00,VMC01} is essentially\footnote{There is some
difference, if state of program register changes in \cite{VC00,VMC01},
even if there is no entanglement in continuous limit under consideration.}
the same as deterministic hybrid gate like \eq{ctrP} discussed in
\Sec{hybproc}.

\end{document}